\documentstyle[12pt]{article}
\input epsf  
\def\bea{\begin{eqnarray}}
\def\eea{\end{eqnarray}}
\begin{document}

\title{Two-gluon form factor of the nucleon and $J/\psi$
photoproduction.}

\author{
L. Frankfurt\\
\it School of Physics and Astronomy, Raymond and Beverly Sackler\\
\it Faculty of Exact Science, Tel Aviv University, Ramat Aviv 69978,\\
\it Tel Aviv , Israel\\
M. Strikman\\
\it Pennsylvania State University, University Park, Pennsylvania 16802}
\date{}
\maketitle
\centerline {\bf ABSTRACT}
We argue that the t-dependence  of the two-gluon form factor
of the nucleon should be
given 
by $\Gamma(t)=(1-t/m_{2g}^2)^{-2}$ with 
$m_{2g}^2\approx 1 GeV^2$. We demonstrate  that this form 
provides a good description of the $t$-dependence of
the cross section of the elastic                
 photoproduction of $J/\psi$-mesons 
between the threshold region of $E_{\gamma}=11~ GeV$ (Cornell), 
$E_{\gamma}=19~ GeV$ (SLAC)
  and  $E_{\gamma}=100~ GeV$ (FNAL) including the
 strong energy dependence of the t-slope.
It is also well matched with the recent HERA data.
The same assumption explains also the t-dependence
of $\phi$-meson electroproduction 
near threshold at $W=2.3~ GeV, Q^2=1.0~ GeV^2$.

\section{Theoretical expectations}
It was demonstrated in \cite{BFGMS} for the case of small x and in \cite{CFS}
 in a general case 
 that in the 
limit of large $Q^2$ the t-dependence of the 
process $\gamma^*_L + N \to V + N$ at fixed $x$ 
is factorized into the convolution of a hard interaction block
calculable in perturbative QCD,  the  short-distance 
$q\bar q $ wave function of the meson, and the generalized/skewed
parton distribution (GPD) in the nucleon.

In the scaling limit the t-dependence is originating solely from the GPD's
since the meson wave function is highly squeezed in the direction transverse to
the reaction axis. In the case of the 
valence quark exchanges these t-dependences are
 constrained by the sum rules for unpolarized 
\cite{sumrules} and polarized quarks \cite{lf} and  could be also
modeled in the chiral soliton models, see review in \cite{GPV}.
 The t-dependence of GPD's provides unique information about the
impact parameter distribution of the partons in nucleons. The 
knowledge of the transverse gluon distribution is
 especially important since it provides a key 
ingredient for the understanding of relative importance 
of soft and hard physics for high energy nucleon-nucleon interactions 
at different impact parameters and has to be implemented in 
 realistic Monte Carlo generators of nucleon-nucleon collisions at collider 
energies.

In this paper we will first discuss 
theoretical expectations for the t-dependence of gluon GPD's.
Next we will use the $11\leq E_{\gamma} \leq 100 GeV$
data to check these  expectations 
and hence to  explain the strong variation with energy of the t-slope
(extracted from the data using $\exp B(t-t_{min})$ fits)
close to the threshold.

First let us summarize the 
theoretical expectations for the t-dependence of the
two gluon form factor:
\begin{equation}
 \Gamma(x_1,x_2,t,\mu^2)={g(x_1,x_2,t,\mu^2)\over g(x_1,x_2,0,\mu^2)},
\end{equation}
where $x_1,x_2$ are the 
light-cone fractions carried by  the 
gluons in the gluon GPD. The fractions 
 are measured relative to the incoming nucleon 
$+$ component, t is usual invariant momentum transfer variable, and
 $\mu^2$ is the renormalization  scale.

If $\mu^2$ is sufficiently large ($\mu^2 \geq 2~ GeV^2$)
 we expect  three 
distinctive regimes of the behavior of $\Gamma$.
At moderate $0.05 \le x_1,x_2\leq 0.3$ the 
form factor should be universal and practically $\mu^2$ independent since it 
is determined by the interaction with the average configurations in the 
nucleon.

The t-dependence in this case could be guessed based on the comparison with 
other nucleon form factors. The difference between the t-dependence
of the electric and 
magnetic form factors  and  the axial form factor is naturally interpreted 
as due to the contribution of the photon scattering off the soft pion cloud,
see e.g. \cite{Weise}. In the case of gluon GPD's 
with $0.05 \le x_1,x_2\leq 0.3$ the pions are not important
since they carry a small 
fractions of the nucleon momentum and also contribute very little 
to the gluon density of the nucleon. Hence a natural guess is that 
\begin{equation}
 \Gamma(x_1,x_2,t,\mu^2)\approx G_A(t), 
\label{axial}
\end{equation}
for $0.05 \le x_1,x_2\leq 0.3$.
In  following analysis we will make {\it a natural  assumption} that
\begin{equation}
 \Gamma(x_1,x_2,t,\mu^2)={1\over (1-t/m_{2g}^2)^2},
\label{dipole}
\end{equation}
for $|t|\le ~ 1\div 2 ~GeV^2$,
with $m^2_{2g}$ expected to be $\sim 1 ~ GeV^2$.
 Similar parameterizations  works well in the case of soft physics
 for the Pomeron coupling with
 the nucleon \cite{DL}. However in this case the mass  scale 
is close to the electromagnetic one, most likely
 because  the soft Pomeron 
interaction with the pion cloud is not suppressed as
 compared to the case of the electromagnetic form factor.

 For significantly smaller  $x_1,x_2$ we expect a 
gradual  increase of the slope \cite{BFGMS},
 which for fixed $x_1-x_2$
 should become
 weaker with increasing $\mu^2$ (as  implicitly discussed
above). In the region of large $x_1,x_2\to 1$ a qualitatively different 
regime is expected when the form factor  should become a very weak function
of x. This is because 
 the transverse momentum is shared between the partons in proportion 
of the longitudinal fractions
(this feature of the Lorentz kinematics plays a critical role in 
 the Feynman mechanism for the nucleon form factor).
 In practice this region  is hard to reach experimentally except 
very close to the photoproduction threshold (see discussion below).

One of the important predictions of the QCD  factorization theorems 
is that for 
 processes dominated by two gluon exchange in the t-channel,
the t-dependence at large $Q^2$ and fixed x
should reach a universal 
limit which is independent of the flavor of the quark constituents 
of the meson \cite{BFGMS}.  The mechanism for such  universality is the 
transverse squeezing of the meson wave function ($r_t\propto 1/Q$). 
Hence in this limit the t-dependence of the amplitude
is given solely by the two-gluon form factor of the nucleon.
The extension of 
the analysis of  \cite{BFGMS} in \cite{FKS96,FKS98} to account for the 
finite transverse size effects for 
$J/\psi$ production the squeezing  starts  already 
from $Q^2\sim 0$. The difference of the 
t-dependences of  $\rho$ and $J/\psi$ production
was calculated in \cite{FKS98} in the dipole approximation.
In that  case, for   $J/\psi$ production the meson 
size contributes $\Delta B \sim 0.3 GeV^{-2}$ at low $Q^2$ and 
 does not change  over experimentally covered range of 
$Q^2$. On the other hand, for   $\rho$ production the slope
strongly depends on $Q^2$.
The  $B_{\rho}(Q^2)- B_{J/\psi}$  difference observed at HERA 
is in  reasonable
 agreement with the \cite{FKS98} prediction for $Q^2 \geq 3~ GeV^2$ 
(see comparison in \cite{Zeuten}).

Another prediction of \cite{BFGMS} was that the rate of 
 change of the $t$-dependence with energy should decrease 
with increase of the hardness of the diffractive process
due to suppression of the 
Gribov diffusion in the hard processes,  at least at moderate x.
The HERA data appear to support this expectation \cite{Levy,H1,ZEUS}.
 If one uses a 
Reggeon type fit one finds for the case of $J/\psi$ production \cite{ZEUS}:
\begin{equation}
B(W)=B_0+2\alpha'\cdot \ln(W/90 GeV)^2,
\end{equation}
where
\begin{eqnarray}
&& B_0=4.30 \pm 0.08(stat)^{+0.16}_{-0.41}(syst) GeV^{-2}  \nonumber \\
&&
\alpha'=0.122\pm 0.033(stat)^{+0.012}_{-0.032}(syst) GeV^{-2}.
\label{al}
\end{eqnarray}
Somewhat smaller values of $\alpha'$ were observed for
 electroproduction of $\rho$-mesons\cite{rhoalpha}.
  $\alpha'$ in Eq. \ref{al} is a  factor 
of $\sim 2$ smaller than $\alpha_{soft}'\approx 0.25~ GeV^2$ measured
for the soft processes.\footnote{A word of caution is in order here. 
At HERA $\alpha'$ was determined by fitting 
$\alpha(t)$ over a large range of $|t|$. 
At the same time a number of measurements of 
elastic hadron-hadron scattering indicate that
$\alpha'_{soft} $ decreases with increase of -t.}
 Moreover  the  $J/\psi$-production 
analysis of \cite{Martin} suggests that the 
observed value of $\alpha'$ can be explained  naturally 
by  the contribution 
of the large size configurations
to
the production  amplitude 
 for which diffusion is not suppressed.

Based on the above discussion 
it appears natural to use the $J/\psi$ photoproduction 
at energies $E_{\gamma}\le ~ 100 GeV$
where Gribov diffusion effects are not important in  the 
extraction of the  two-gluon form factor of the nucleon.

The important effect which we encounter here is that the
t-dependence  of the cross section which follows from Eq.~\ref{dipole},
\begin{equation}
{d\sigma \over dt}\propto  \Gamma^2(x_1,x_2,t,\mu^2)={1\over (1-t/m_{2g}^2)^4},
\label{dipole2}
\end{equation}
 does not exactly match an  exponential form. As a result we expect that the effective exponential  slope
would depend on the $t$-interval used in the data analysis.
This is especially true for low energies where $-t_{min}$ is not equal to zero.

If  defines the  slope as the logarithmic derivative of the cross section,
then  the  fit to Eq. \ref{dipole2} becomes:
\begin{equation}
B_{eff}(t)={4\over m_{2g}^2-t}
\label{sleff}
\end{equation}
The slope of the exponential fit, $B$, corresponds roughly to $B_{eff}$
calculated for the average t of the experiment. In the case of HERA data
for the lowest end of the energy interval  the data give 
$B\sim 3 ~GeV^{-2}$. We checked that this corresponds to 
\begin{equation}
m_{2g}\approx 1 GeV.
\label{m2g}
\end{equation}
Since we neglected in Eq. \ref{dipole2}  the contribution
of the $J/ \psi$ size which contributes 
$\Delta B \approx 0.3 GeV^{-2}$ to the slope, we expect the
value of $m^2_{2g}$ should be larger than the result of our fit by about 0.1 
GeV$^{-2}$.

\section{Comparison with the data}
Let us now check the consistency of Eqs.~\ref{dipole2},\ref{m2g}
 with the data 
obtained at lower energies. We want to emphasize here that we do not
 feel that the accuracy of the data and information available about the
systematics justifies at that stage performing a $\chi^2$ fit. We simple fix 
the value of $m_{2g}$ and check whether a reasonable description can be
 achieved.

First we consider the FNAL data of Binkley et al. \cite{Binkley}
at $\left<E_{\gamma}\right>=100 ~ GeV$
which  appear to be
the only data where the recoil proton was detected. In other measurements
the elastic sample was 
contaminated (especially at large $|t|$) by
 the inelastic diffractive events.
The comparison of the t-dependence is presented in Fig. 1. One can see 
that a good agreement with the shape of these data is reached.

Next we consider the photoproduction data at energies close to the threshold.
In this case $t_{min}$ is not negligible. If the form factor t-dependence
 is indeed a power law it would result in  a decrease of the slope
of the exponential fits of the form $\exp B(t-t_{min})$. Two experiments
 reported the t-dependence in this range. The SLAC data \cite{Camerini}
 at $ E_{\gamma}=19~ GeV$
correspond to $-t_{min}=0.087~ GeV^2$. We find them in a good agreement
with  Eq. \ref{dipole2}, see Fig. 2.  
The Cornell experiment \cite{Gittelman}
 measured $J/\psi $ photoproduction at
 $\left< E_{\gamma}\right>=11 ~GeV$. Because 
 $-t_{min}=0.41 GeV^2$, 
 we expect a  significant  change of the slope. The data presented in Fig. 3
indeed correspond to a very weak t-dependence. \footnote{The paper 
\cite{Gittelman} in addition to the plot of the data presents also a 
fit to the data $B=1.25\pm 0.2$ which corresponds to a significantly
slower t-dependence than indicated by the plot. 
Reasons for this are not clear.}
 It is probably somewhat weaker than the 
expectation based on   Eq.\ref{dipole2}. However one should 
remember that in this case the value of longitudinal light-cone fraction 
transfered from the nucleon to $J/\psi$: $x_1-x_2=1-p_{f, N}^+/p_{i, N}^+$
 becomes  large (it is equal to 0.48 for $\left< E_{\gamma}\right>=11~ GeV$).
\footnote{Often $x_1-x_2$ for $J/\psi$ photoproduction is calculated as 
$m_{J/\psi}^2/s$. 
This expression is not valid very close to the threshold.
However it does a good 
job even for $ E_{\gamma}=11~ GeV$. } It was suggested in \cite{Brodsky}
that photoproduction of charm near threshold is dominated by a 
three gluon exchange with the nucleon rather than by a two -gluon exchange as at high energies.  The observed connection of the $t$-dependence of
$J/\psi$ production at 11 GeV and at higher energies indicates that at least for 11 GeV the dominant contribution remains a two gluon exchange.

Hence we conclude that the current data are in a 
reasonable agreement with the suggested form Eq. ~\ref{dipole2}. Clearly new 
much more accurate data close to threshold are necessary. The planned 
SLAC experiment E160
 \cite{E160} may contribute here as well as the Jlab 12 GeV upgrade.
Another critical test will be a measurement of the $t$-dependence of the 
$\Upsilon$ photoproduction at HERA which should be very close to
 the genuine two-gluon form factor. Note in passing that 
the t-slope of the elastic
$\Upsilon$ photoproduction entered in the calculation of
the total cross section of the $\gamma + p\to \Upsilon +p$ 
reaction in 
\cite{martinupsilon}. Since at that time the data on the energy dependence of the
slope of $J/\psi$ production did not allow determination of
$\alpha'$ for $J/\psi$ we took it to be equal to zero.
Taking  $\alpha'$ from Eq. \ref{al} leads to renormalization of our 
prediction in \cite{martinupsilon}  by a 
factor $\sim 1.3$.

It is natural also to ask a question whether a large part of the 
variation of the t-slope of the exclusive
electroproduction of light vector mesons near threshold could  
be due to a similar effect. The cleanest case is the $\phi-$meson production
for which the quark exchange contribution is strongly suppressed.
Naturally in this case the size of the meson cannot be as safely
 neglected as in the $J/\psi$ case. At the same time  the threshold region
corresponds to large values of $x_1-x_2\approx 0.4$ which work
 in the direction of slowing
down the t-dependence of the cross section. Hence we performed a comparison of 
 Eq. \ref{dipole2} with the most recent data on 
electroproduction of  $\phi-$meson which were reported in the Jlab experiment
 \cite{Jlab} (the Cornell data \cite{Cornell} are pretty similar). 
Surprisingly enough we find  reasonable agreement with the data
- Fig.4.
This suggest a common origin of the dynamics of the $\phi$-meson
and $J/\psi$
electroproduction  near threshold.

In conclusion, we have demonstrated that the current data 
are consistent with the dipole dependence of the two-gluon form factor 
with the mass scale $m^2_{2g}\approx 1~ GeV^2$, which is a lower bound 
due to the contribution of the finite  $J/\psi$ size which is of the order 
$0.1 ~GeV^2$.
 This corresponds to a  significantly 
smaller radius of the distribution of the 
gluon field in the nucleon than for 
the electromagnetic charge where $m^2\sim ~ 0.7 GeV^2$. It is 
 close to the mass scale in the axial form factor which reflects
the distribution of the valence quarks. 
A more narrow space distribution at $x\ge 0.05$, especially 
when combined with a small value of $\alpha'$ for virtualities $\ge 2 
~GeV^2$,
has many implications for  diffractive studies as well as for modeling
the structure of final states at high energies in pp collisions with 
high $p_t$ jets.

We thank  K.~Griffioen for useful comments and GIF and DOE for support.

\begin{figure}
\begin{center}
        \leavevmode
        \epsfxsize=.80\hsize
       \epsfbox{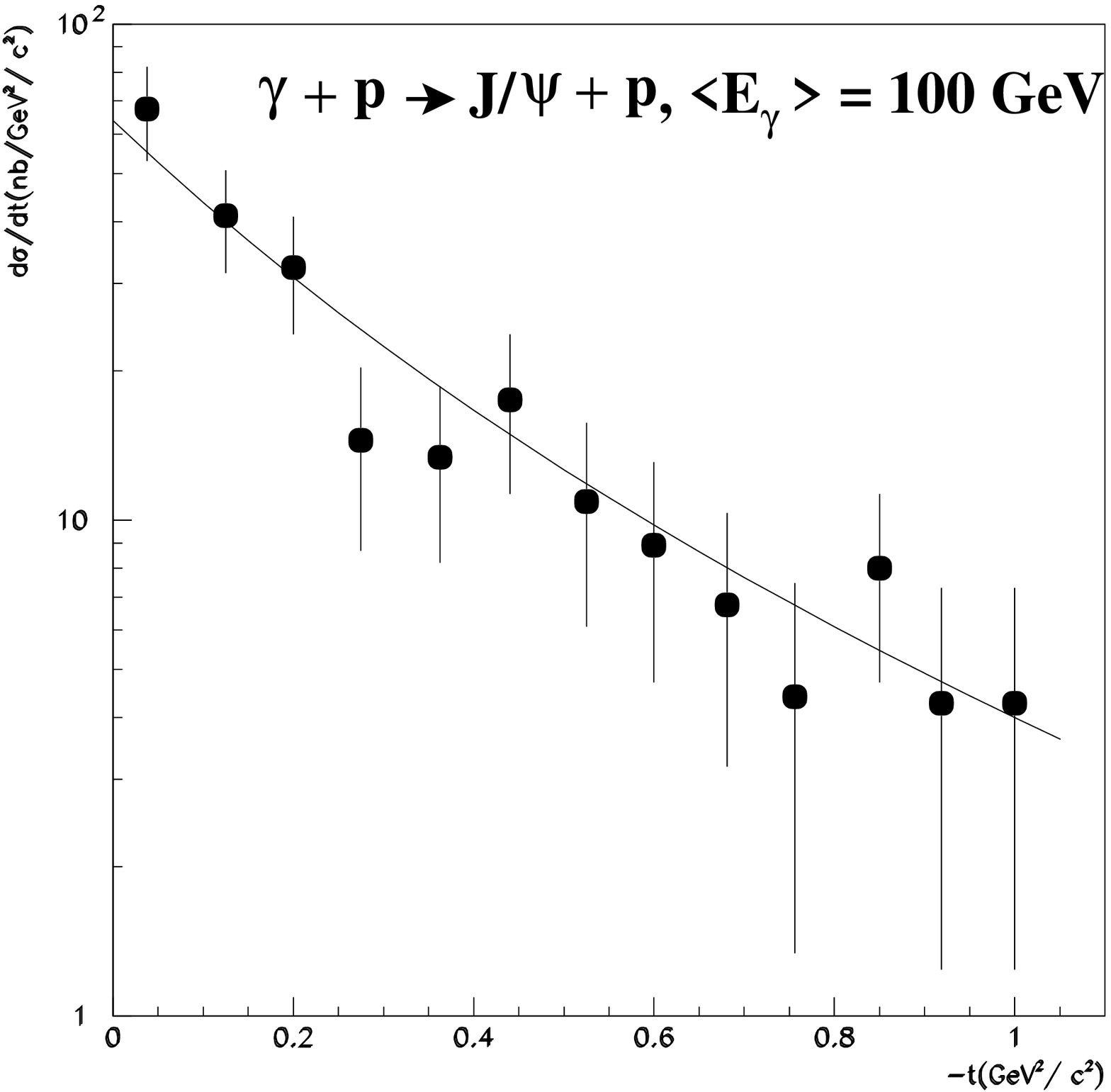}
    \end{center}
\vspace*{1cm}
\caption{
Comparison of the dipole parameterization of the $d\sigma^{\gamma +p \to J/ \psi +p}/dt$ 
with the data of \cite{Binkley} at $\left<E_{\gamma}\right>=100~ GeV$.
}
\label{100}
\end{figure}
\begin{figure}
\begin{center}
        \leavevmode
        \epsfxsize=.80\hsize
       \epsfbox{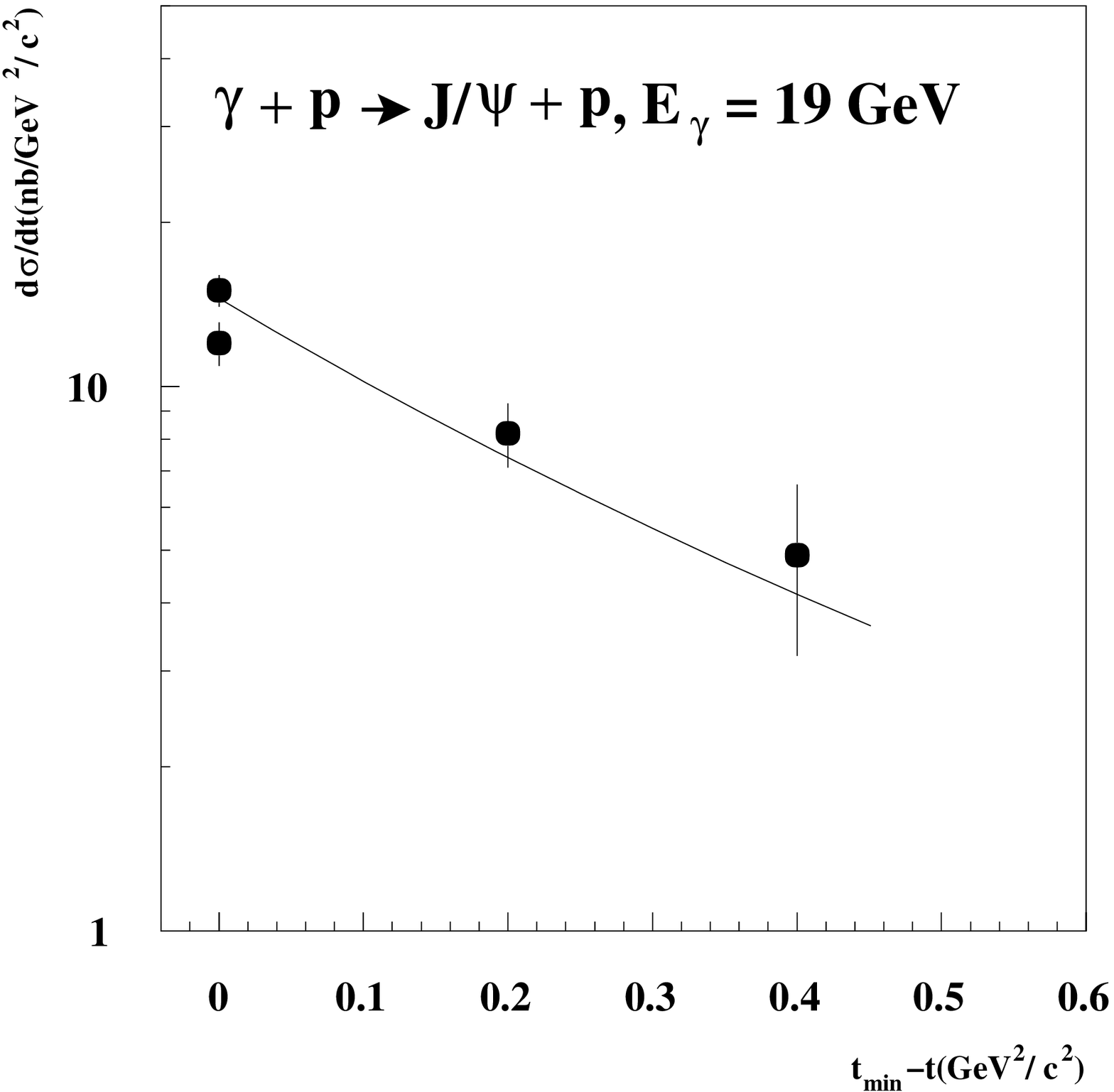}
    \end{center}
\vspace*{1cm}
\caption{
Comparison of the dipole parameterization of the $d\sigma^{\gamma +p \to J/ \psi +p}/dt$ 
with the data of \cite{Camerini} at $E_{\gamma}=19~ GeV$
}
\label{slacfig}
\end{figure}
\begin{figure}
\begin{center}
        \leavevmode
        \epsfxsize=.80\hsize
       \epsfbox{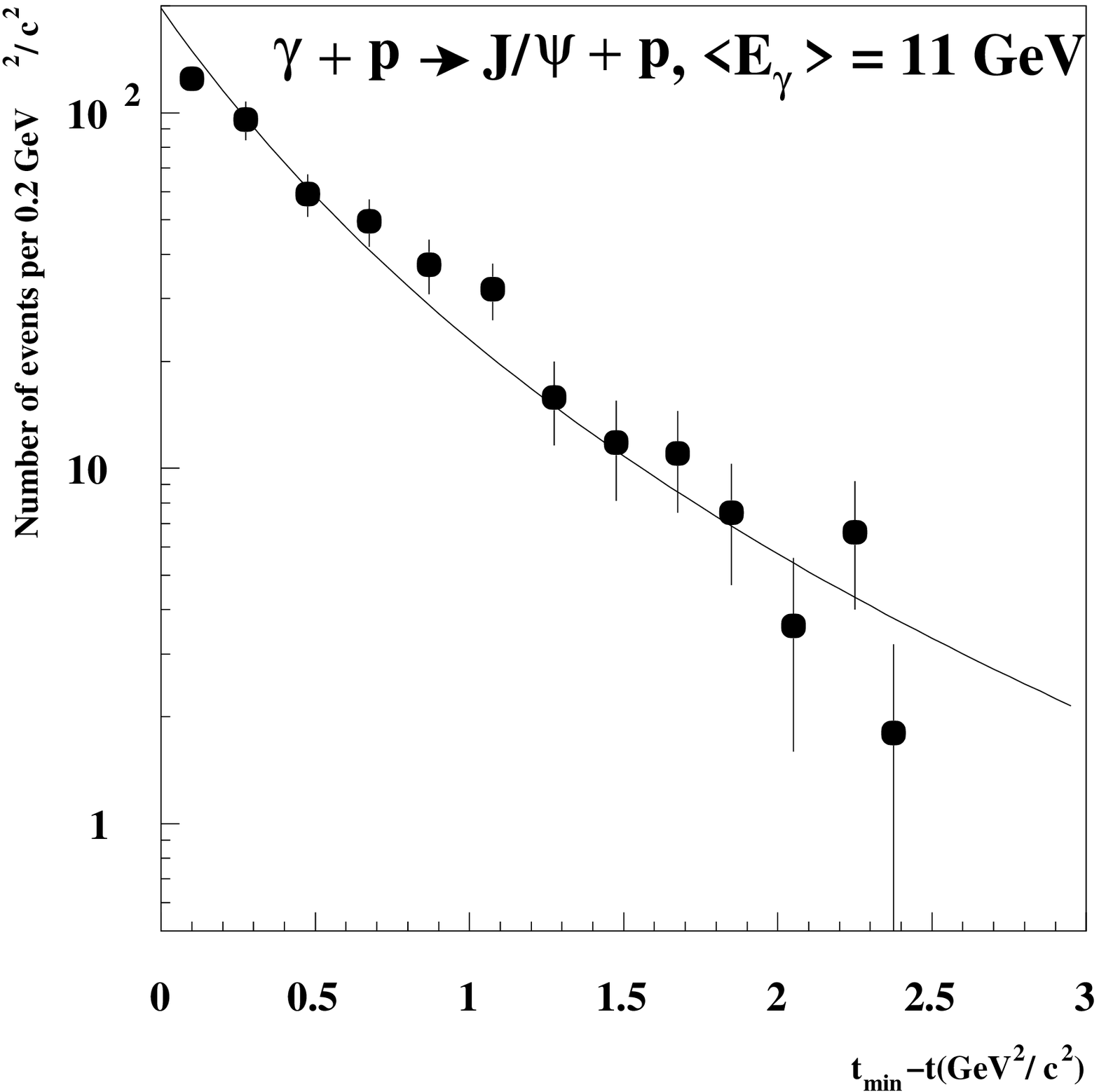}
    \end{center}
\vspace*{1cm}
\caption{
Comparison of the dipole parameterization of the $d\sigma^{\gamma +p \to J/ \psi +p}/dt$ 
with the data of \cite{Gittelman} at $\left<E_{\gamma}\right>=11 ~GeV$.
}
\label{cornfig}
\end{figure}
\begin{figure}
\begin{center}
        \leavevmode
        \epsfxsize=.80\hsize
       \epsfbox{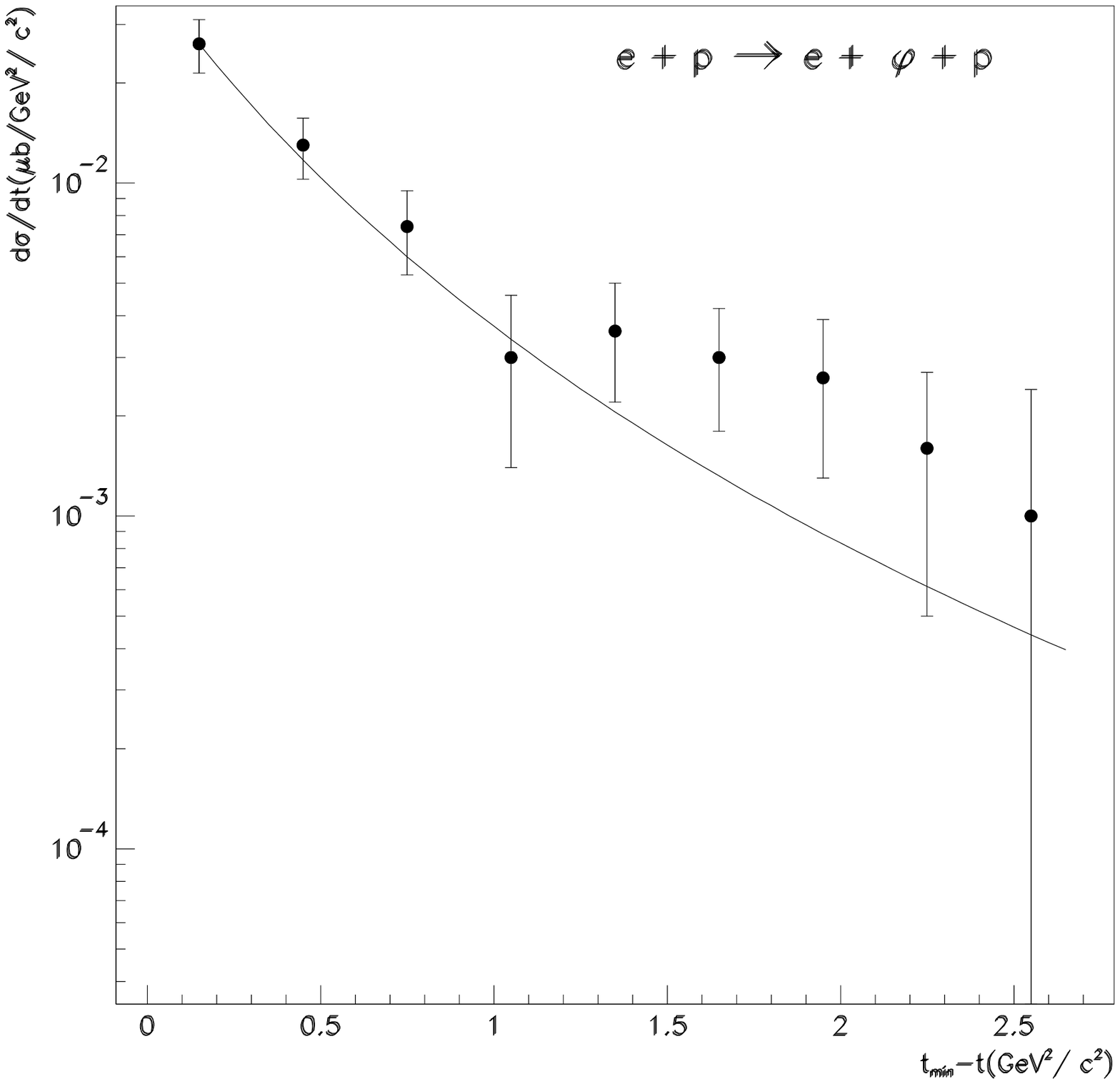}
    \end{center}
\vspace*{1cm}
\caption{
Comparison of the dipole parameterization of the $d\sigma^{\gamma^* +p \to \phi +p}/dt$ 
with the data of \cite{Jlab} at $\left<W\right>=2.3 ~GeV, 
\left<Q^2\right>=1 ~GeV^2$.
}
\label{phifig}
\end{figure}

\begin{thebibliography}{99}


\bibitem{BFGMS}
S.~J.~Brodsky, L.~Frankfurt, J.~F.~Gunion, A.~H.~Mueller and M.~Strikman,
Phys.\ Rev.\ D {\bf 50}, 3134 (1994)
[arXiv:hep-ph/9402283].
\bibitem{CFS}J.~C.~Collins, L.~Frankfurt and M.~Strikman,
Phys.\ Rev.\ D {\bf 56}, 2982 (1997)
[arXiv:hep-ph/9611433].
\bibitem{sumrules}
V. De Alfaro, S. Fubini, G. Furlan, C. Rossetti, Currents in hadron physics,
1973.

\bibitem{lf}
L.~L.~Frankfurt,
Yad.\ Fiz.\  {\bf 16}, 1250 (1972).

\bibitem{GPV}
K.~Goeke, ~V.~Polyakov and M.~Vanderhaeghen,
Prog.\ Part.\ Nucl.\ Phys.\  {\bf 47}, 401 (2001)
[arXiv:hep-ph/0106012].
\bibitem{Weise}
E.~Oset, R.~Tegen and W.~Weise,
Nucl.\ Phys.\ A {\bf 426}, 456 (1984)
[Erratum-ibid.\ A {\bf 453}, 751 (1986)].
\bibitem{FKS96}
L.~Frankfurt, W.~Koepf and M.~Strikman,
Phys.\ Rev.\ D {\bf 54}, 3194 (1996)
[arXiv:hep-ph/9509311].
\bibitem{FKS98}
L.~Frankfurt, W.~Koepf and M.~Strikman,
Phys.\ Rev.\ D {\bf 57}, 512 (1998)
[arXiv:hep-ph/9702216].

\bibitem{Zeuten}
L.~Frankfurt and M.~Strikman,
Nucl.\ Phys.\ Proc.\ Suppl.\  {\bf 79}, 671 (1999)
[arXiv:hep-ph/9907221].


\bibitem{Levy}
A.~Levy,
Phys.\ Lett.\ B {\bf 424}, 191 (1998)
[arXiv:hep-ph/9712519].

\bibitem{H1}
C.~Adloff {\it et al.}  [H1 Collaboration],
Phys.\ Lett.\ B {\bf 483}, 23 (2000)
[arXiv:hep-ex/0003020].
\bibitem{ZEUS}
S.~Chekanov {\it et al.}  [ZEUS Collaboration],
arXiv:hep-ex/0201043.
\bibitem{rhoalpha}A.Kowal, ZEUS contribution to the DIS2001 on vector 
meson production.
\bibitem{Martin}
L.~Frankfurt, M.~McDermott and M.~Strikman,
JHEP {\bf 0103}, 045 (2001)
[arXiv:hep-ph/0009086].


\bibitem{DL}
A.~Donnachie and P.~V.~Landshoff,
Nucl.\ Phys.\ B {\bf 267}, 690 (1986).

\bibitem{Binkley}
M.~Binkley {\it et al.},
Phys.\ Rev.\ Lett.\  {\bf 48}, 73 (1982).


\bibitem{Camerini}
U.~Camerini {\it et al.},
Phys.\ Rev.\ Lett.\  {\bf 35}, 483 (1975).


\bibitem{Gittelman}
B.~Gittelman, K.~M.~Hanson, D.~Larson, E.~Loh, A.~Silverman and G.~Theodosiou,
Phys.\ Rev.\ Lett.\  {\bf 35}, 1616 (1975).

\bibitem{Brodsky}
S.~J.~Brodsky, E.~Chudakov, P.~Hoyer and J.~M.~Laget,
Phys.\ Lett.\ B {\bf 498}, 23 (2001)
[arXiv:hep-ph/0010343].


\bibitem{E160} V.Ghazikhanian, et al, 
SLAC proposal E160; http://www.slac.stanford.edu/exp/e160.

 \bibitem{martinupsilon}L.~L.~Frankfurt, M.~F.~McDermott and M.~Strikman,
JHEP {\bf 9902}, 002 (1999)
[arXiv:hep-ph/9812316].


\bibitem{Jlab}
K.~Lukashin {\it et al.}  [CLAS Collaboration],
Phys.\ Rev.\ C {\bf 63}, 065205 (2001)
[arXiv:hep-ex/0101030].

\bibitem{Cornell}
R.~Dixon {\it et al.},
Phys.\ Rev.\ D {\bf 19}, 3185 (1979).

\end{thebibliography}
\end{document}